\documentclass[a4paper, 12pt,apl]{revtex4-2}

\usepackage{amssymb}
\usepackage{amsmath}
\usepackage{epsfig}
\usepackage{graphicx}
\usepackage{multirow}
\usepackage{xcolor}

\DeclareMathOperator{\Tr}{Tr}

\begin{document}

\author{R. J. Hudspith}
\email{renwick.james.hudspith@googlemail.com}
\affiliation{PRISMA$+$ Cluster of Excellence and Institut für Kernphysik,
  Johannes Gutenberg-Universit\"at Mainz, 55099 Mainz, Germany}
\author{D. Mohler}
\email{d.mohler@gsi.de}
\affiliation{GSI Helmholtzzentrum f\"ur Schwerionenforschung, 64291 Darmstadt, Germany}

\title{A fully non-perturbative charm-quark tuning\\
 using machine learning}

\begin{abstract}
We present a relativistic heavy-quark action tuning for the charm sector on ensembles generated by the CLS consortium. We tune a particular 5-parameter action in an entirely
non-perturbative and -- up to the chosen experimental input --
model-independent way using machine learning and the continuum experimental charmonium
ground-state masses with various quantum numbers. In the end we are reasonably
successful; obtaining a set of simulation parameters that we then verify
produces the expected spectrum. In the future, we will use this action for finite-volume
calculations of hadron-hadron scattering. 
\end{abstract}
\maketitle
\newpage

\section{Introduction}

%Since the dawn of time man has often looked up into the sky and wondered how to tune the Fermilab charm quark action in a fully-nonperturbative model-independent way.

In recent years, studies of hadron-hadron scattering using L\"uscher's finite
volume method \cite{Luscher:1986pf,Luscher:1990ux} have become the state of the art for obtaining
information about physical scattering amplitudes from Lattice QCD. In the
light-quark sector, such studies typically make use of moving frames to maximise the information about
scattering amplitudes deduced from a single gauge field ensemble (for examples, please refer to \cite{Briceno:2017max}).

For hadrons with heavy charm or bottom quarks, using the Wilson-clover action
\cite{Sheikholeslami:1985ij} employed for the light (up, down, and strange) quarks leads to large
heavy-quark discretisation effects. These discretisation effects can be
understood in the context of the Fermilab approach \cite{El-Khadra:1996wdx,Oktay:2008ex}. While the
simplest implementation of such a relativistic heavy quark (RHQ) action used by the Fermilab Lattice and MILC
collaborations (see \cite{FermilabLattice:2010rur,FermilabLattice:2014ysv}) tunes the kinetic masses of
spin-averaged ground states and thereby obtains an excellent description of
ground-state mass splittings in the charmonium spectrum \cite{DeTar:2018uko}, large discretisation effects are still
present in hadronic dispersion relations \cite{Oktay:2008ex,Lang:2014yfa}.

For the study of hadron resonances and bound states with L\"uscher's finite
volume method, such an approach was used for heavy-light mesons in \cite{Mohler:2013rwa,Lang:2014yfa,Lang:2015hza} and for charmonium in \cite{Lang:2015sba}. Due to the
non-standard dispersion relation in the Fermilab method with
\cite{El-Khadra:1996wdx,Oktay:2008ex} $M_1\ne M_2\ne M_4\dots$, these studies
only used rest-frame data, as there is no simple expression for a change of
frame from interacting two-hadron energies in moving frames to the center-of-momentum frame consistent with the Fermilab method
dispersion relation. In more recent studies of the charmonium spectrum in (coupled-channel) scattering of
charmed mesons \cite{Piemonte:2019cbi,Prelovsek:2020eiw}, moving frames were
used by only extracting energy differences with regard to close-by non-interacting meson-meson levels and by purely
working with the continuum form of the dispersion relations, which minimises the effects from
heavy-quark discretisation effects
% on the dispersion relations
\footnote{For a
  more thorough description of this procedure see Section IV B. of \cite{Piemonte:2019cbi}}.
This introduces a hard-to-quantify systematic uncertainty.

To reduce the uncertainty from
heavy-quark discretisation effects in the context of scattering studies, it is therefore desirable to at least
tune the dispersion relation to be approximately relativistic with a speed of
light $c=1$ (in natural units). In the next section we will describe the lattice gauge field
ensembles used in our simulation, introduce the relativistic heavy-quark
action used, and describe the methodology for a fully non-perturbative tuning
of this action. In Section \ref{results} we proceed to show the outcome of
this tuning procedure, and compare it to the charm simulations with the CLS
light-quark action. In Section \ref{outlook} we summarize our results and
provide a brief outlook.

\section{Methodology}

\subsection{Gauge field ensembles}

\begin{table}[tbh]
\begin{tabular}{cc|ccc|cc}
\toprule
$\beta$ & Name & $L^3\times L_T$ & T-Boundary & $N_\text{conf}\times N_\text{src}$ & $a^{-1}$ [GeV] & $m_\pi$ [GeV] \\
\hline
3.34 & A653 & $24^3\times 48$  & Periodic & $100\times 16$ & 1.987(20) & 0.422(4) \\
3.34 & A654 & $24^3\times 48$  & Periodic & $100\times 16$ & 1.987(20) & 0.331(3) \\
\hline
3.40 & U103 & $24^3\times 128$ & Open     & $1000\times 1$ & 2.285(28) & 0.419(5) \\
3.40 & H101 & $32^3\times 96$  & Open     & $500\times 1$  & 2.285(28) & 0.416(5) \\
3.40 & H102 & $32^3\times 96$  & Open     & $500\times 1$  & 2.285(28) & 0.354(5) \\
3.40 & H105 & $32^3\times 96$  & Open     & $500\times 1$  & 2.285(28) & 0.284(4) \\
\hline
3.46 & B450 & $32^3\times 64$  & Periodic & $100\times 16$ & 2.585(33) & 0.416(4) \\
3.46 & S400 & $32^3\times 128$ & Open     & $500\times 1$  & 2.585(33) & 0.351(4) \\
3.46 & N451 & $48^3\times 128$ & Periodic & $253\times 8$  & 2.585(33) & 0.287(4) \\
\hline
3.55 & H200 & $32^3\times 96$  & Open     & $1000\times 1$ & 3.071(36) & 0.419(5) \\
3.55 & N202 & $48^3\times 128$ & Open     & $900\times 1$  & 3.071(36) & 0.410(5) \\
3.55 & N203 & $48^3\times 128$ & Open     & $750\times 1$  & 3.071(36) & 0.345(4) \\
3.55 & N200 & $48^3\times 128$ & Open     & $856\times 1$  & 3.071(36) & 0.282(3) \\
\hline
3.70 & N300 & $48^3\times 128$ & Open     & $770\times 1$  & 3.962(45) & 0.421(4) \\
\botrule
\end{tabular}
\caption{Table of ensembles, values for the lattice spacing come from \cite{Gerardin:2019rua}
  and \cite{Chao:2021tvp}, as do the pion masses. Further details on these ensembles can be
  found in \cite{Bruno:2014jqa}. Here, $N_\text{conf}$ indicates the number of well-separated gauge configurations used and $N_\text{src}$ the number of time-translated wall-sources averaged over per gauge configuration.}
\label{configtable}
\end{table}

We perform calculations on gauge-field ensembles generated by the Coordinated Lattice Simulations (CLS)
consortium \cite{Bruno:2014jqa,Bali:2016umi} with 2+1 flavours of dynamical, non-perturbatively improved Wilson
fermions \cite{Sheikholeslami:1985ij}. We use 5 different lattice spacings ranging from $a=0.09929$ to $0.04981$ fm, and with pion masses between 421 and 282 MeV, on trajectories where the trace
of the quark mass matrix $\Tr{M}$ is kept constant and approximately
physical. In our study we use ensembles with either open \cite{Luscher:2012av} or periodic boundary
conditions in time. The ensembles we considered for the tuning of
the relativistic heavy-quark action are listed in Tab.~\ref{configtable}. 

\subsection{Charm-quark action}

We will follow \cite{Aoki:2003dg}, also known in the literature as the ``Tsukuba'' action, and write down a general asymmetric Wilson action,
\begin{align}\label{eq:tsuk_action}
  D_{xy}=\delta_{xy}&-\kappa_c\bigg[\sum_i(r_s-\nu\gamma_i)U_i(x)\delta_{x+i,y}+(r_s+\nu\gamma_i)U_i^\dagger(x)\delta_{x,y+i}\bigg]\nonumber\\
                    &-\kappa_c\bigg[(r_t-\gamma_t)U_t(x)\delta_{x+t,y}+(r_t+\gamma_t)U_t^\dagger(x)\delta_{x,y+t}\bigg]\\
  &-\kappa_c\bigg[c_B\sum_{i,j}\sigma_{ij}F_{ij}(x)+c_E\sum_i\sigma_{it}F_{it}(x)\bigg]\delta_{xy}.\nonumber
\end{align}
Commonly, $r_t=1$ is chosen as this parameter is argued to be redundant. The remaining five free parameters are $\kappa_c$, $r_S,\nu,c_E,$ and $c_B$. Prescriptions in the literature exist for choosing these parameters based on
mean-field-improved perturbation theory and for non-perturbative tuning of
individual parameters. % and other such mysticism.
An in-depth discussion about our implementation of this action within the library \verb|openQCD| can be found in App.~\ref{sec:charm_impl}.

In \cite{FermilabLattice:2010rur,FermilabLattice:2014ysv},
$c_{sw}=c_E=c_B\equiv u_0^{-3}$ is set to the value from tree-level
tadpole-improved perturbation theory, with the tadpole factor $u_0$ determined
from the fourth root of the plaquette, or the mean Landau link. In this
approach only $\kappa_c$ is determined non-perturbatively (from tuning the
kinetic mass $M_2$ of the $D_s$-meson to its physical value), making predictions of the resulting charmonium spectrum and splittings.

In heavy-hadron spectroscopy studies pursued by the Hadron Spectrum Collaboration, anisotropic
lattices are used to study multi-hadron scattering with heavy hadrons
\cite{Moir:2016srx,Cheung:2016bym,Cheung:2017tnt,Cheung:2020mql,Gayer:2021xzv}. In
their approach the fine temporal lattice spacing combined with a tuning of
the mass and anisotropy parameters that reproduce the physical $\eta_c$ mass
and a relativistic dispersion relation enables the use of
moving frames. Discretisation effects in the spin-dependent
splittings are still, however, expected to be sizable in such an approach \cite{Oktay:2008ex}.

In \cite{Aoki:2003dg}, the full 5-parameter action of Eq.~\ref{eq:tsuk_action} was introduced. In \cite{PACS-CS:2011ngu,PACS-CS:2013vie}\footnote{And similarly for \cite{Kayaba:2006cg} except for the tuning in $\nu$.} this action was \textit{semi} non-perturbatively tuned on a set of ensembles with the same lattice spacing $a^{-1}=2.194(10)\text{ GeV}$ where $\kappa_c$ was tuned to obtain the physical spin-average of the $\eta_c$ and $J/\psi$ masses. The parameter $r_s$ was determined completely perturbatively using one-loop mean-field improved expressions, whereas $c_E$ and $c_B$ were determined from perturbative mass-corrections to the non-perturbatively measured $c_{\text{SW}}$ of \cite{CP-PACS:2005igb}. The value of $\nu$ was tuned independently of all the other parameters to obtain the relativistic dispersion relation $c^2=1$. We will discuss the validity of such an approach later in Sec.~\ref{sec:comp_varied_nu}, as we allow all parameters to vary and can infer their inter-dependence. Having $c^2=1$ is quite useful as we will not have to determine masses of states through their dispersion relation; in the Fermilab language we are enforcing that $M_1=M_2$.

Finally, in \cite{Lin:2006ur} a more conventional non-perturbative tuning of the
4-parameter action with $r_t=1$ and $r_s=\nu$, and of the 3-parameter action
where additionally $c_E=c_B$ has been performed. As we see no reason why these
parameters would be independent, in fact we will see they are related in a
somewhat complicated manner, we prefer to follow a much more global and general approach.

We will follow App.~A of \cite{Hudspith:2020tdf} (in that context used for
tuning an NRQCD action) and use partially-twisted Coulomb-gauge-fixed wall
sources \footnote{Fixed to an accuracy of $10^{-14}$ using the CFACG algorithm
  \cite{Hudspith:2014oja} implemented in GLU.} to determine the spin-averaged
dispersion relation of the $\eta_c$ and $J/\psi$. We will partially-twist
\cite{Bedaque:2004kc,Sachrajda:2004mi} along the diagonal
$(\theta,\theta,\theta)$ to maximally-reduce effects from rotational symmetry
breaking and use the following 5 twist angles (in lattice units):
\begin{equation}
\theta_i = 0,\;\sqrt{\frac{2}{9}},\;\sqrt{\frac{4}{9}},\;\sqrt{\frac{6}{9}},\;\sqrt{\frac{8}{9}}, \nonumber
\end{equation}
such that our values of $(ap)^2$ are evenly-distributed between 0 and 3.

\begin{figure}[tb]
\includegraphics[scale=0.35]{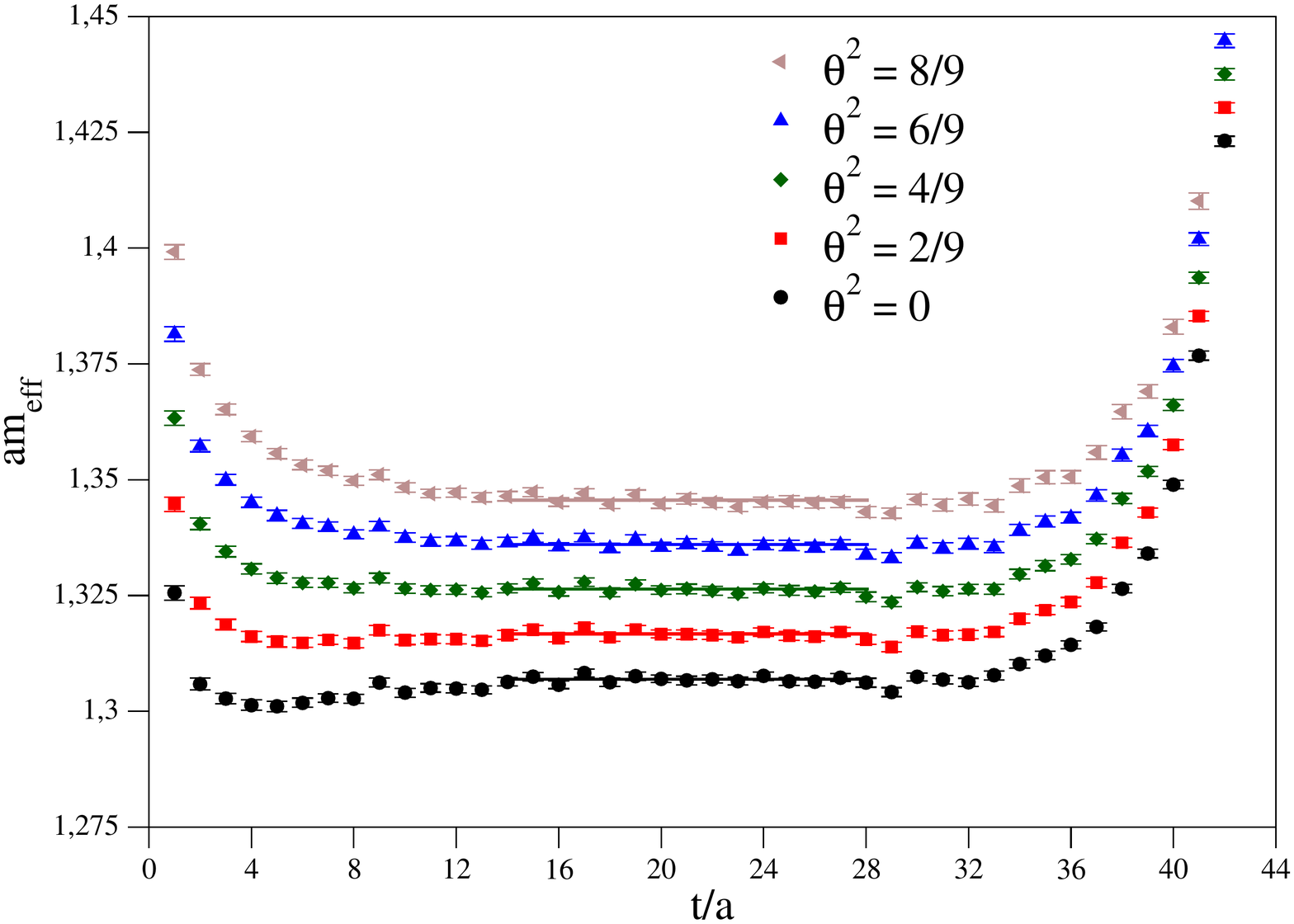}
\caption{Effective mass plots of the $\eta_c$-meson with the fit of Eq.~\ref{eq:disp} for the ensemble H101.}\label{fig:H101_etac}
\end{figure}

To determine $c^2$ and our masses $am$ we will fit the $\eta_c$ and $J/\psi$ correlators to the following form,
\begin{equation}\label{eq:disp}
C(p,t) = A(1+p^2(D+p^2\;E))\left(e^{-t\sqrt{(am)^2+c^2(ap)^2}}+Be^{-(L_T-t)\sqrt{(am)^2+c^2(ap)^2}}\right),
\end{equation}
with free parameters $A,am,D,E$ and $c^2$ the speed of light squared. The
parameter B takes the value of 1 if the gauge field has a periodic temporal
boundary and 0 if it is open. An example of our tuning and the fit to
Eq.~\ref{eq:disp} for the $\eta_c$ correlator on the ensemble H101 is shown in Fig.~\ref{fig:H101_etac}.

\subsection{Chosen measurements and correlation functions}

\begin{figure}[tb]
\includegraphics[scale=0.35]{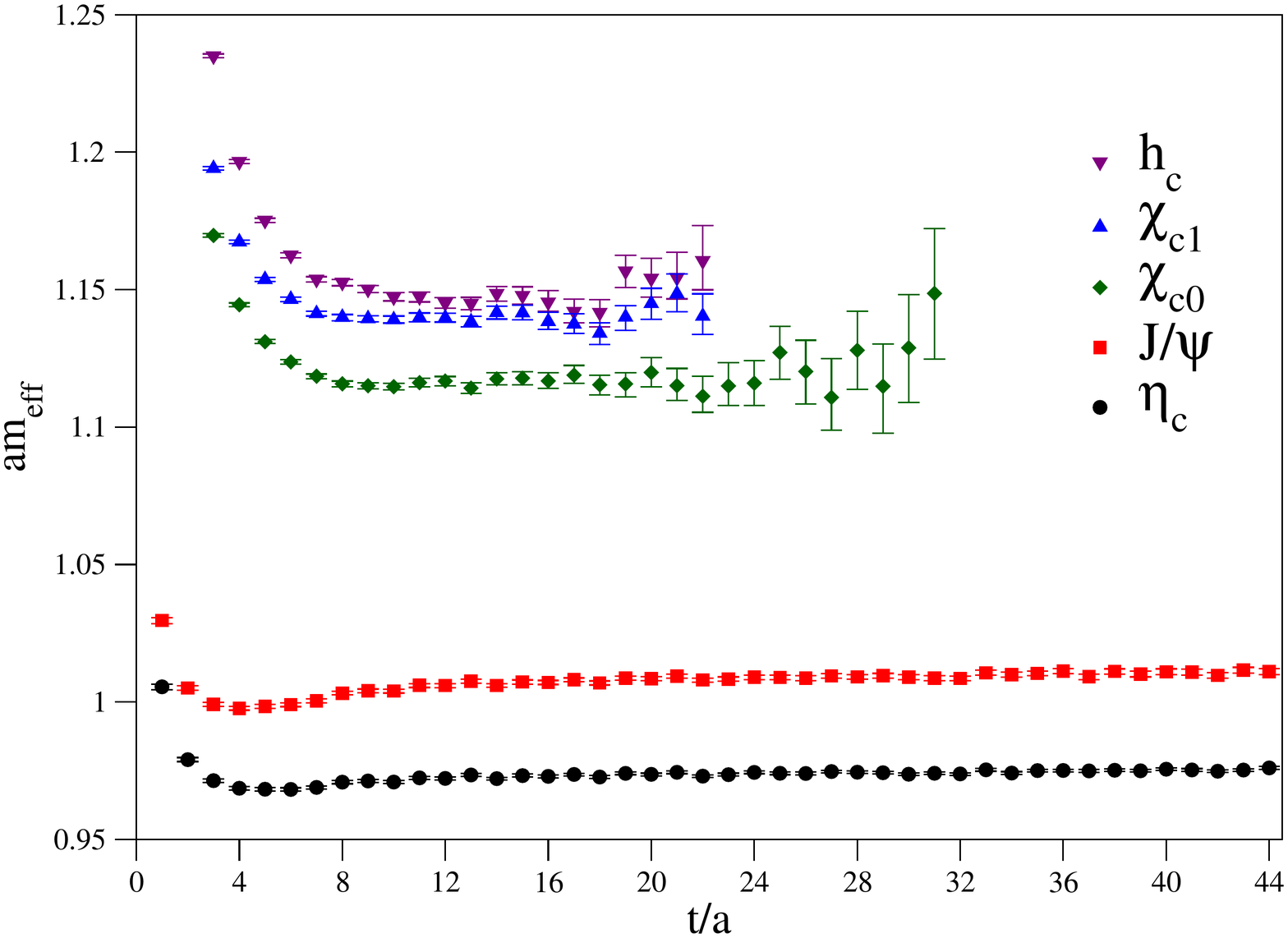}
\caption{Effective masses of the states listed in Tab.~\ref{tab:states} for the ensemble N202.}\label{fig:N202_masses}
\end{figure}

To make sure that we are using observables that are sensitive to all of the
parameters in the RHQ action, we will be using both S- and P-wave charmonium
ground-state masses (listed in Tab.~\ref{tab:states}) in addition to the dispersion relation for the spin-averaged
S-wave ground state. A possibility that would make this dependence a bit more
obvious would have been to focus on the 1S
hyperfine splitting, along with the S-wave -- P-wave splitting, the spin-orbit
splitting and the tensor splitting within the P-wave, as for example used in
\cite{Burch:2009az,DeTar:2018uko}. For computational reasons, we instead
restrict the interpolator basis to the states that can be
reached using the simple mesonic operators at rest,
\begin{equation}
O(x) = (\bar\psi \Gamma \psi)(x),
\end{equation}
and compare these directly with their continuum analogs. Note that the
$J^{PC}=1^{+-}$ $h_c$ is however a good proxy for the spin-average of the 1P
states, as the 1P hyperfine splitting is expected to be very small
\cite{Lebed:2017yme} (which can also be seen in a previous lattice determination \cite{DeTar:2018uko}). Throughout, we will use gauge-fixed wall sources with
Gaussian sink-smearing implemented by the method discussed in
App.~\ref{app:bsink} to optimise for the ground states and those with twisting. An example of our zero-momentum states can be found in Fig.~\ref{fig:N202_masses} for the ensemble N202.

\begin{table}
  \begin{tabular}{c|ccccc}
    \toprule
    State & $\eta_c$ & $J/\psi$ & $\chi_{c0}$ & $\chi_{c1}$ & $h_{c}$ \\
    \hline
    $\Gamma$ & $\gamma_5$ & $\gamma_i$ & $I$ & $\gamma_i\gamma_t$ & $\gamma_i\gamma_j$ \\
    $J^{PC}$ & $0^{-+}$ & $1^{--}$ & $0^{++}$ & $1^{++}$ & $1^{+-}$ \\
    Experiment [GeV] & 2.9839 & 3.096916 & 3.41471 & 3.51072 & 3.52549 \\    
    \botrule
  \end{tabular}
  \caption{List of operators used in our measurement of charmonium, their expected quantum number equivalents to continuum states and the experimental masses of these states \cite{Zyla:2020zbs}.}\label{tab:states}
\end{table}

For the open-boundary ensembles we perform a single wall-source propagator inversion at $T/2$, in the middle of the bulk, and symmetrise the resulting correlator. We must make sure to not perform our fits too close to the open boundary as these boundary-effects can be relatively strong, as seen in Fig.~\ref{fig:H101_etac}.

\subsection{Neural-network charm-action tuning}

Our idea is to randomly choose values of $\kappa_c,r_s,\nu,c_E,$ and $c_B$ and
measure the basic spectrum of Tab.~\ref{tab:states} on a sufficient number of gauge configurations. We will then train a
neural network in predicting these parameters for given input masses and $c^2$, and
finally use this model to predict the best guesses for the parameters of the
action that lie closest to the experimental masses with spin-averaged $c^2=1$, hopefully within a precision of $1\%$. We acknowledge that all of the states will likely have residual cut-off
effects with different slopes but the parameters of the action should be able
to absorb the leading cut-off effects \cite{El-Khadra:1996wdx,Oktay:2008ex}. As we approach the (chiral-)continuum limit the states from the predicted parameters tend to the continuum masses, by design.

We strived to have at least about 30 different sets of run parameters randomly drawn
from initially broad Gaussian distributions per ensemble \footnote{To start with, we draw values of $c_E$ and $c_B$ around $c_SW$ with a bias towards larger values and a large width, as motivated by perturbation theory. $\kappa_c,r_s,$ and $\nu$ are initially guessed from a linear fit in $a^2$ from previously-measured ensembles.}. For some ensembles we performed more runs to investigate the
possible inter-ensemble dependencies, such as those emanating from the pion
mass or the volume. Typically, after about 20 runs we start refining our
parameter space by narrowing the new trial parameters and by feeding guesses back into the system to accelerate convergence. For each state entering the
network we use 1000 bootstraps, which will both help give the network lots of data to train against and inform it of correlations between states.

We will first consider each individual ensemble separately to train the
network, eventually investigating combinations of ensembles and performing
more-global fits. We initially have 6 input parameters: the lattice
determinations of the $\eta_c,J/\psi,\chi_{c0},\chi_{c1},$ and $h_c$ mesons,
and the lattice $\eta_c-J/\psi$ spin-averaged dispersion relation $c^2$. We
will use a feed-forward network with one hidden layer of only 12
neurons \footnote{We tried many different numbers of layers and neurons in each layer. Enlarging these numbers either did not change predictions compared to the small, simple network, or resulted in what we determined to be over-fitting.} and an output layer of 5 neurons ($\kappa_c,r_s,\nu,c_E,$ and $c_B$). On each of the layers we use the \verb|sigmoid| activation and the \verb|Adam| minimiser, with an adjusted learning rate, early-stopping, and a batch-size of 80. With the loss-function set as the mean-squared deviation. Often the model will converge rapidly to stable training and validation losses in fewer than 200 epochs. 

Once the network \textit{understands} the relation between the lattice-determined states and the charm-action parameters on an ensemble, we determine the best finite-lattice-spacing approximation to the action parameters needed to obtain something close to the experimental states with $c^2=1$ from the spin-averaged dispersion relation. Initial studies indicated that this approach gave predictions with heavier-than-physical $\eta_c$ and $\chi_{c0}$ states, and we decided to introduce so-called "class weights" to tell the network to prefer tuning $\kappa_c$ by a factor of 2 compared to the other outputs. Empirically, this makes the $\eta_c$ and $\chi_{c0}$ closer to physical while making the $\chi_{c1}$ and $h_c$ and $c^2$ a little less accurate, as will be seen in Fig.~\ref{test_spect}.

For the selection of our training and validation sets we randomly shuffle the
runs and remove approximately $20\%$ for validation. We run the training over
50 of these reshufflings and perform a weighted average (with weight
proportional to the inverse of the validation loss) of the networks' best
guesses to determine our predictions. In the following (Tab.~\ref{tab:prediction_pars_run04}) we quote our predicted parameters with $1\sigma$ errors, although this should not be construed as a true error but rather an indication of the variation of the parameters within the training runs we performed.

\section{Results\label{results}}

\subsection{Table of predictions}

\begin{table}[h]
\begin{tabular}{c|c|ccccc}
\toprule
Ensemble & Runs & $\kappa_c$ & $c_E$ & $c_B$ & $r_s$ & $\nu$ \\
\hline
A653 & 41 & 0.10951(09) & 2.150(16) & 2.313(28) & 1.1645(34) & 1.1515(40) \\
A654 & 33 & 0.10939(06) & 2.110(11) & 2.290(13) & 1.1709(39) & 1.1609(24) \\
\hline
Comb & 74 & 0.10946(04) & 2.138(09) & 2.305(13) & 1.1672(23) & 1.1563(25) \\
\hline
\hline
U103 & 40 & 0.11409(09) & 1.982(06) & 2.139(15) & 1.1375(13) & 1.1229(16) \\
H101 & 40 & 0.11436(10) & 1.925(11) & 2.112(18) & 1.1412(22) & 1.1222(16) \\
H102 & 45 & 0.11447(07) & 1.894(12) & 2.032(16) & 1.1454(20) & 1.1098(28) \\
H105 & 40 & 0.11450(09) & 1.937(08) & 2.093(09) & 1.1376(22) & 1.1144(19) \\
\hline
Comb & 165& 0.11434(06) & 1.940(07) & 2.102(12) & 1.1395(18) & 1.1167(23) \\
\hline
\hline
B450 & 34 & 0.11826(10) & 1.890(08) & 2.051(08) & 1.1017(23) & 1.0849(36) \\
S400 & 35 & 0.11796(09) & 1.880(10) & 2.060(12) & 1.1073(17) & 1.0924(19) \\  
N451 & 23 & 0.11835(10) & 1.907(09) & 2.025(07) & 1.1020(15) & 1.0950(20) \\
\hline
Comb & 92 & 0.11810(04) & 1.894(04) & 2.052(05) & 1.1043(12) & 1.0926(12) \\
\hline
\hline
H200 & 36 & 0.12248(11) & 1.864(09) & 1.894(07) & 1.0683(22) & 1.0518(29) \\
N202 & 36 & 0.12256(20) & 1.865(08) & 1.889(10) & 1.0679(13) & 1.0446(24) \\
N203 & 36 & 0.12232(06) & 1.863(09) & 1.879(11) & 1.0733(09) & 1.0605(11) \\
N200 & 36 & 0.12211(09) & 1.860(08) & 1.888(06) & 1.0753(09) & 1.0564(15) \\
\hline
Comb & 144& 0.12235(05) & 1.859(05) &  1.886(04) & 1.0722(09) & 1.0552(08) \\
\hline
\hline
N300 & 50 & 0.12608(15) & 1.792(15) & 1.873(14) & 1.0382(13) & 1.0168(19) \\
\botrule
\end{tabular}
\caption{Charm-action parameter predictions for our data, the quantities in
  brackets are one-sigma variations in the predicted parameters, see text for
  details. "Comb" refers to the prediction from the neural-network when
  combining all results with a given lattice spacing. Different lattice
  spacings are separated by double horizontal lines.}\label{tab:prediction_pars_run04}
\end{table}

In Tab.~\ref{tab:prediction_pars_run04} we present our results for the tuning
on individual ensembles as well as a combined data-set built from the
individual ones at fixed lattice spacing, which is practically an assumption
of a flat approach to physical light and strange quark masses, combined with
the (likely justified) assumption that finite volume effects in our charmonium
observables are not relevant for our volumes. Typically the loss from the neural network reduces when the larger data-set "Comb" is used, and this larger data-set should protect us a little more from over-fitting.

Comparing our results to the similar tuning of \cite{PACS-CS:2011ngu}
($\kappa_c=0.10959947$, $c_E=1.7819512$, $c_B=1.9849139$, $r_s=1.1881607$, and
$\nu=1.1450511$) albeit on different ensembles, we see some similarities and a
few differences. Namely, that at a similar lattice spacing to theirs
($a^{-1}=2.194\text{ GeV}$), our 1-ensembles, the values we obtain for
$\kappa_c,c_E$, and $c_B$ are larger, while both $r_s$ and $\nu$ are smaller. Our
larger values of $c_E$ and $c_B$ are unsurprising as the gauge fields used are
generated with the tree-level Symanzik gauge action whereas they used the
Iwasaki, and the size of the non-perturbative $c_{\text{SW}}$ is larger for the Symanzik gauge action at comparable lattice spacing. It is interesting to note that their values for $\kappa_c,r_s$, and $\nu$ resemble more our tuning parameters on the coarser ensembles A653 and A654.

It is clear from Tab.~\ref{tab:prediction_pars_run04} that as the
lattice-spacing reduces the size of all of the parameters $c_E,c_B,r_s,$ and
$\nu$ also decrease. With $r_s$ and $\nu$  practically tending to their expected continuum values of 1. Both of the parameters $c_E$ and $c_B$ are greater than the value of (but tending toward) $c_\text{SW}$ from \cite{Bulava:2013cta}, with the hierarchy of $c_B>c_E$ as expected perturbatively from \cite{Aoki:2003dg}.

\subsection{On the validity of fixing parameters and tuning others}\label{sec:comp_varied_nu}

As $c^2=1$ from the spin-averaged dispersion relation is one of our inputs we can use this as a handle and investigate the dependence on the predicted \textit{physical} parameters as this value is varied (and all the other states are still set to their physical values). If the deviation on $\kappa_c,r_s,c_E,c_B,$ and $\nu$ is large as small changes are applied to this dispersion relation, then the model is suggesting that there is significant inter-dependence between these parameters and that this tuning shouldn't be done independently, as was performed in \cite{PACS-CS:2011ngu}.

\begin{figure}[tb]
  \includegraphics[scale=0.35]{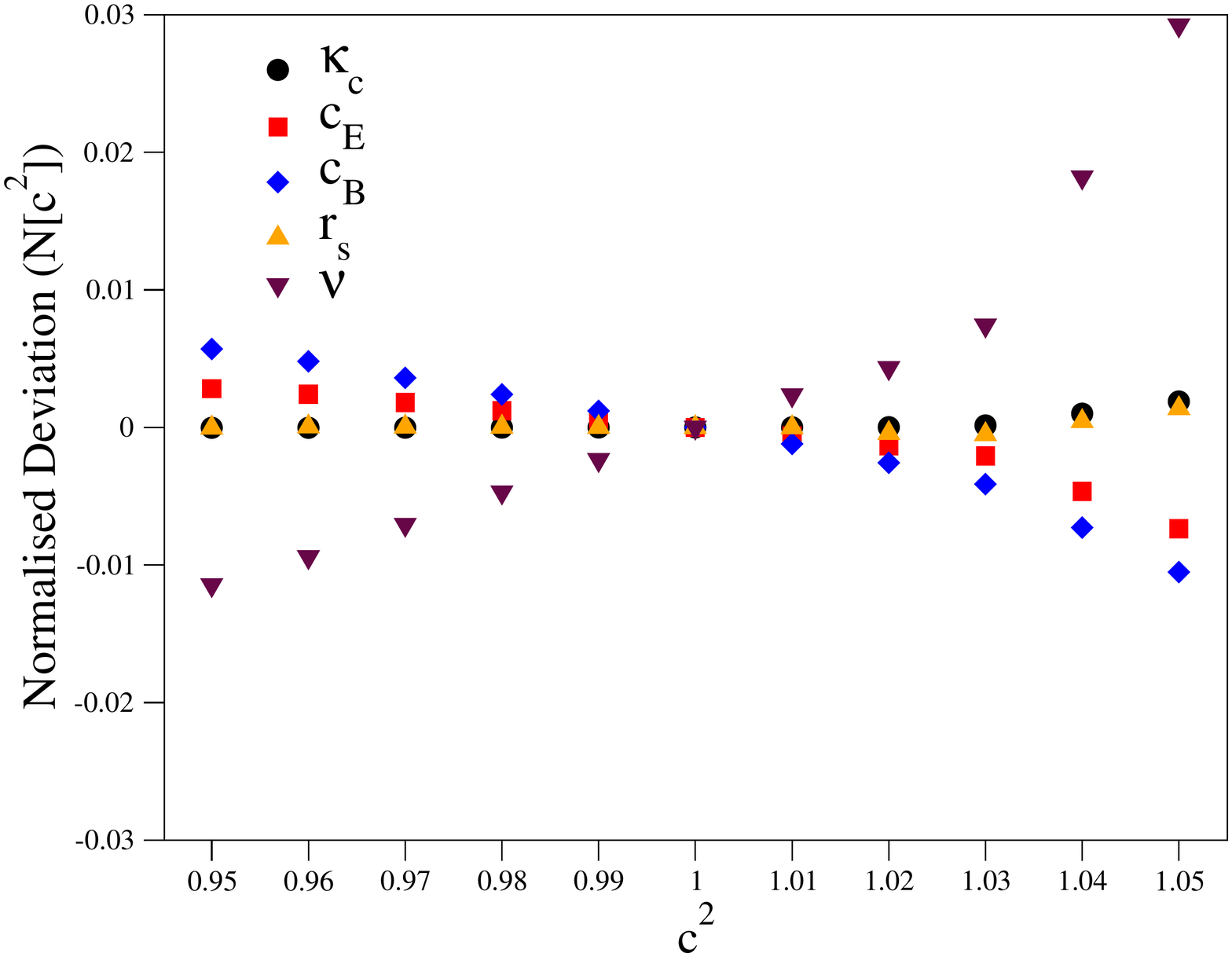}
  \caption{The predicted normalised deviation of the parameters as we change the effective spin-averaged speed of light.}\label{fig:cdep}
\end{figure}

Fig.~\ref{fig:cdep} illustrates the normalised deviation for some model-predicted parameter P, which we define as,
\begin{equation}
N[c^2] = (P[c^2]-P[1])/P[1],
\end{equation}
for the combined $A653$ and $A654$ ensembles as we varied the spin-averaged
dispersion relation's $c^2$. We can see that $\kappa_c$ and $r_s$ do not depend strongly on $c^2$ whereas $\nu,c_E,$ and $c_B$ do. With $c_E$ and $c_B$ being somewhat anti-correlated with $\nu$. Although $\nu$ seems to be the main perpetrator in determining the effective speed of light, $c_E$ and $c_B$ both change accordingly and would likely need to be re-tuned to accommodate an independent tuning in $\nu$.

\subsection{Accuracy of our predictions and systematics}

\begin{figure}[p]
\includegraphics[width=0.848\textwidth]{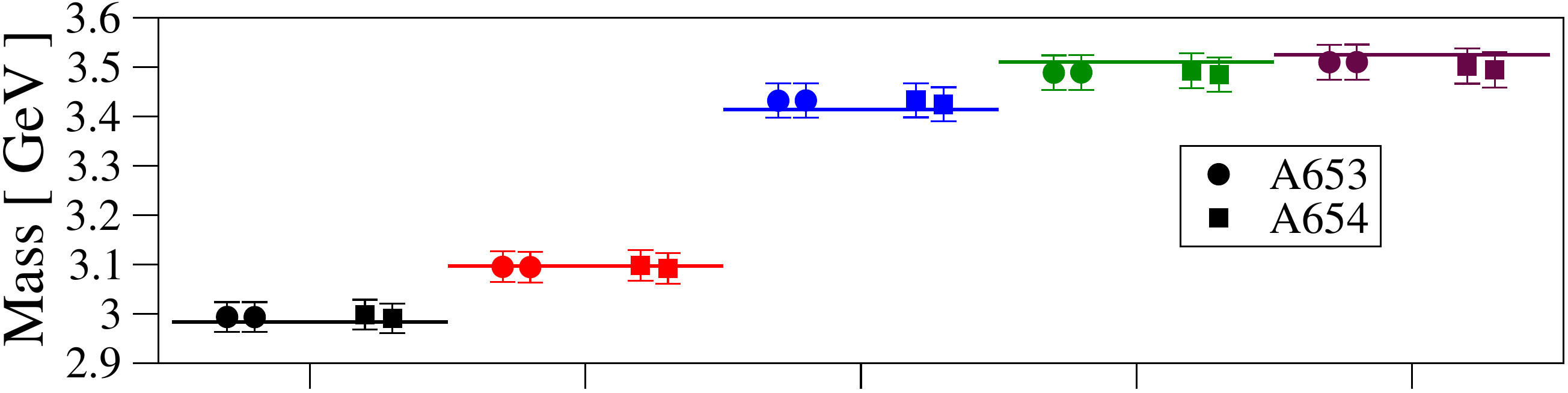}
\includegraphics[width=0.848\textwidth]{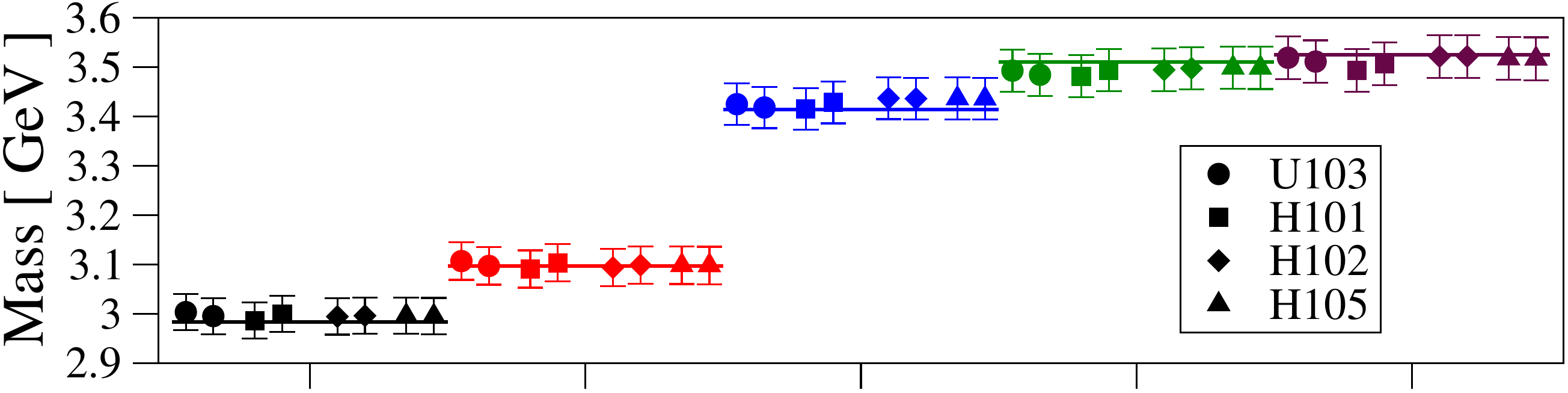}
\includegraphics[width=0.848\textwidth]{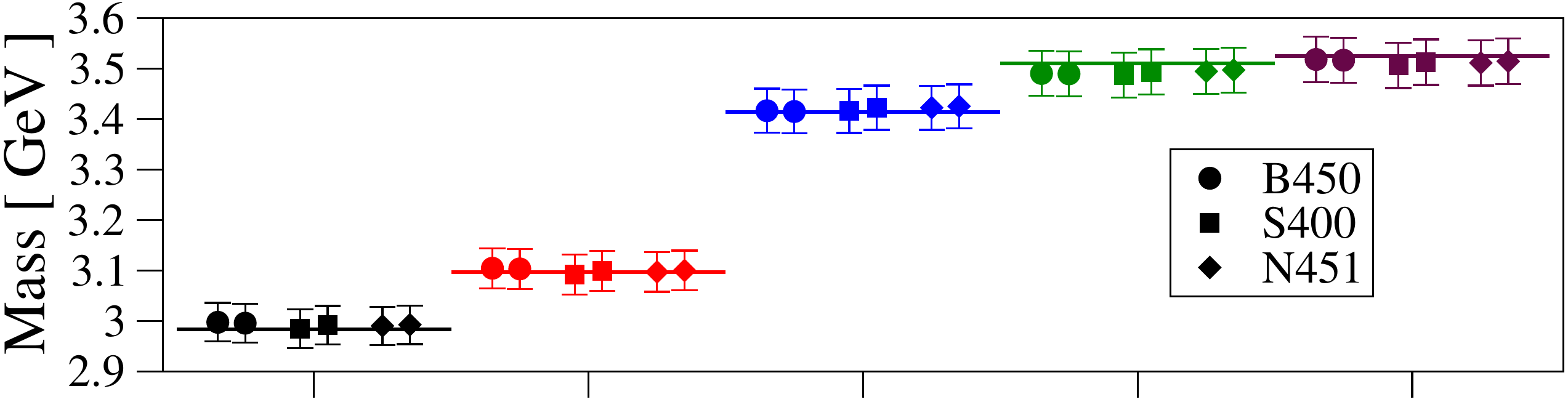}
\includegraphics[width=0.848\textwidth]{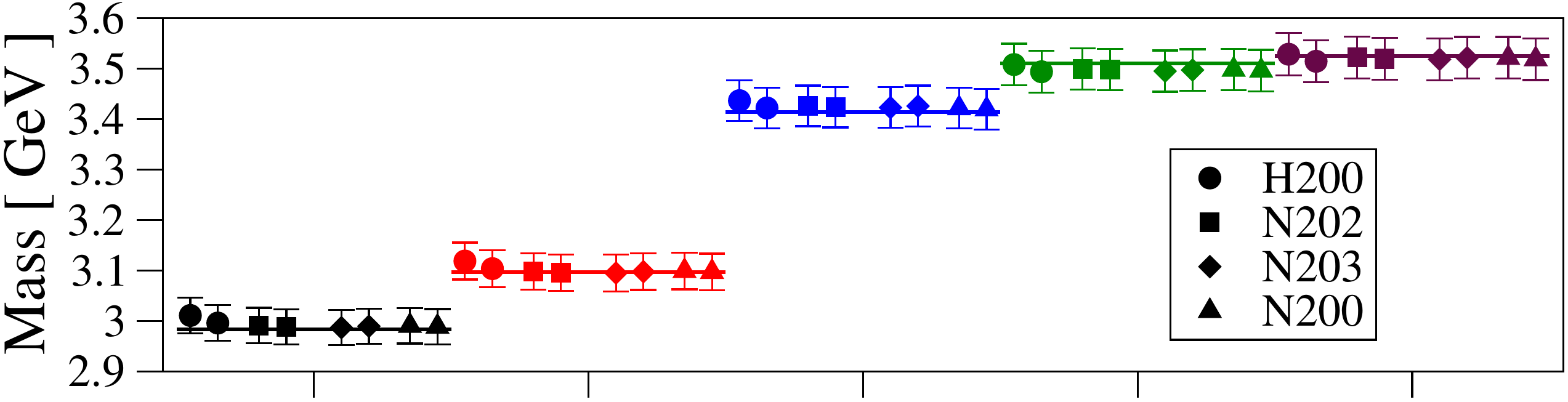}
\includegraphics[width=0.848\textwidth]{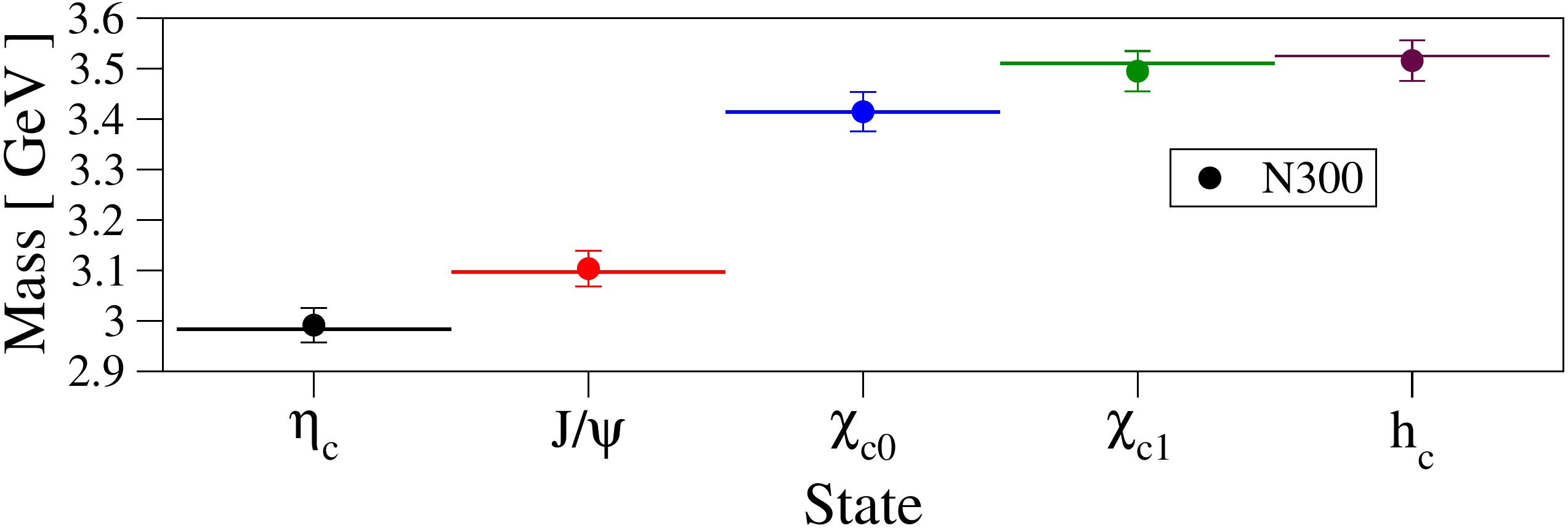}
\caption{Overview of the resulting masses for the hadrons used for our neural
  net tuning. The horizontal lines show the experiment masses. The uncertainty combines statistical and scale-setting
  uncertainties in quadrature, and the uncertainty is dominated by the scale
  setting. The panes are sorted from coarsest~(top) to finest~(bottom)
  lattice spacing. For each ensemble two results are shown: the tuning on the single ensemble
  (left symbol) and the tuning on the collection of ensembles with the same
  $\beta$ (right symbol). Ensembles within the same panel differ by volume and/or pion mass.}\label{test_spect}
\end{figure}

We now proceed to test our predicted parameters from Tab.~\ref{tab:prediction_pars_run04} by performing the same measurement runs on the same configurations as the training was performed on, and comparing the determined states and the effective speed of light to the physical continuum states. We will investigate both the combined and the individual results to illustrate the lack of measurable dependence on the pion mass within the precision of our determinations. A plot of our ability to reproduce the spectrum is shown in Fig.~\ref{test_spect} and one can see that within our large errors there is good agreement, mostly this is due to the lack of precision of our lattice spacing.

\begin{figure}[tb]
\includegraphics[width=0.6\textwidth]{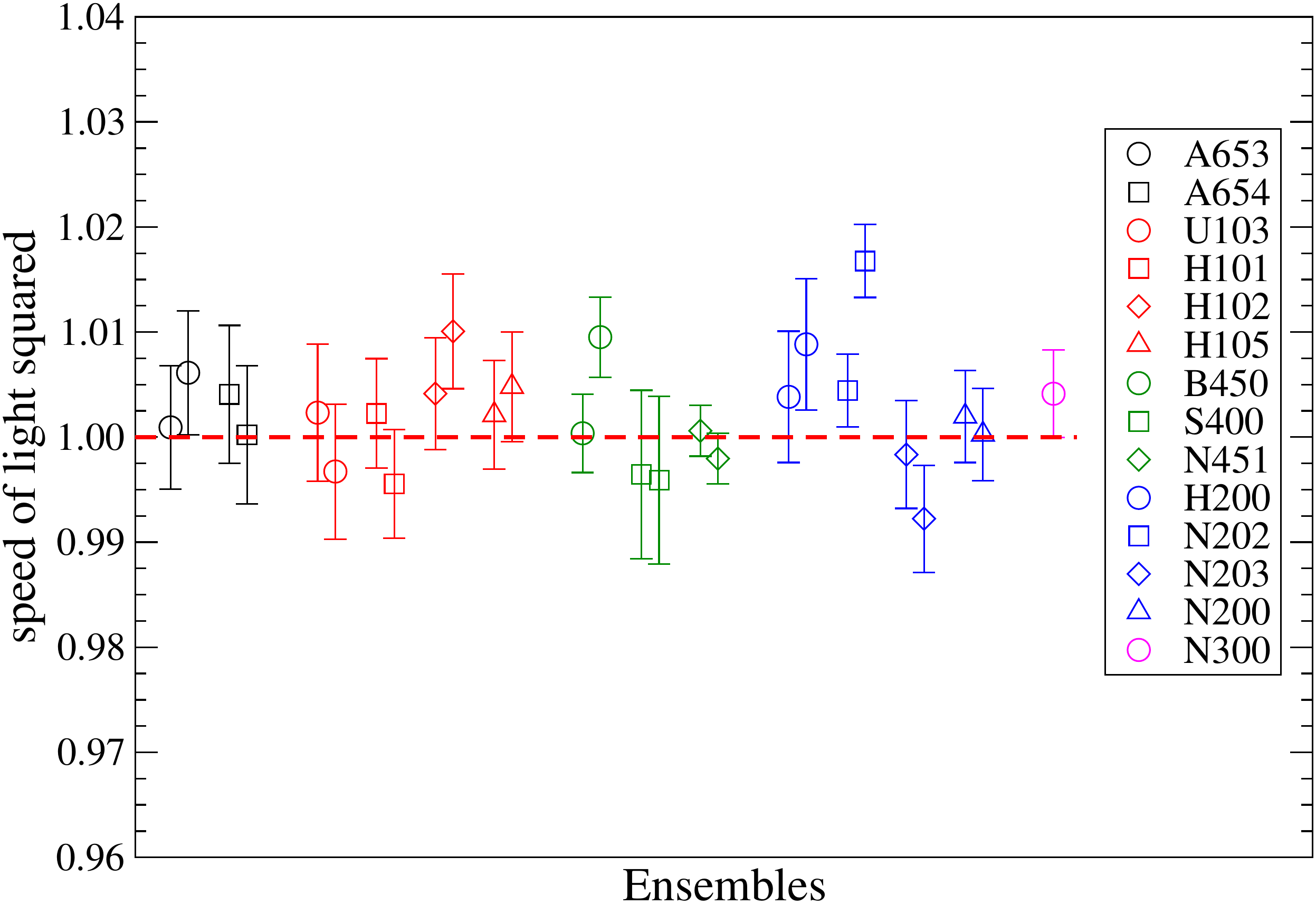}
\caption{Spin-averaged $c^2$ reached through our tuning
  prescription. For each ensemble a pair of data points is shown. The left
  data point result from the tuning on the single
  ensemble, while the right data point results from the tuning
  using all ensembles with the same lattice spacing.}\label{csquared}
\end{figure}

Fig.~\ref{csquared} shows the spin-averaged $c^2$ reached for our tuning procedure
on all ensembles. For each ensemble we show both the result from the
tuning procedure on a single ensemble (the left point of the pair) and the result from using all ensembles
at the same value of $\beta$ (the right point of a given pair). Our overall goal here was to to obtain $c^2=1$
to about 1\%, which we achieve on most ensembles. For the dispersion relation,
the single-ensemble tuning works a bit better than combining ensembles at each
lattice spacing. However, the opposite will be true for the splitting
discussed in the next few paragraphs. Note, that while there are some outliers,
the situation is vastly improved compared to a standard Wilson-action
($c_B=c_E=c_{sw}$, and $r_s=r_t=\nu=1$) as used for the charm quark in \cite{Gerardin:2019rua}, and discussed in the next sub-section (Sec.~\ref{Sec:stupid_tune}). 

\begin{figure}[tb]
\includegraphics[width=0.48\textwidth]{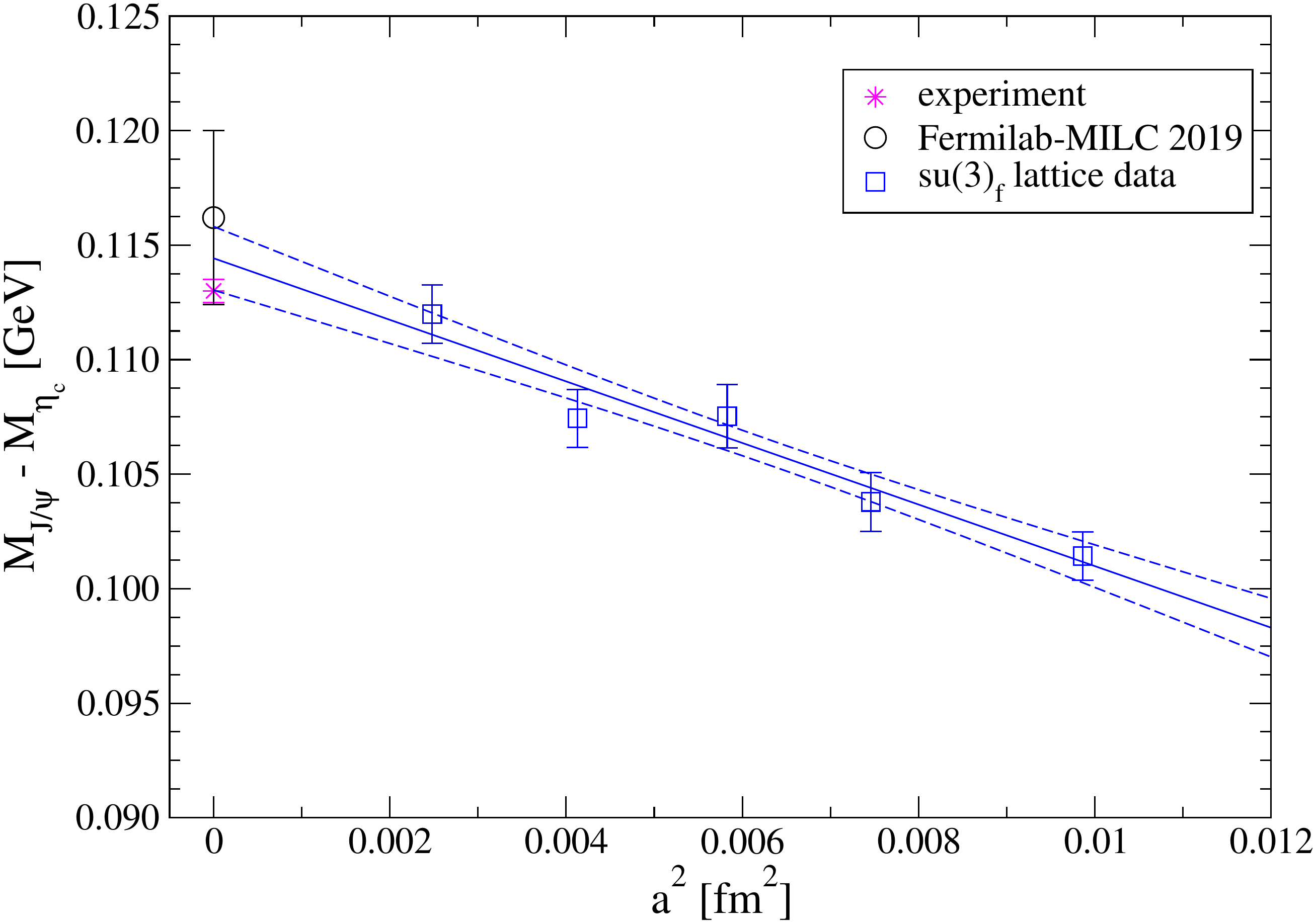}
\includegraphics[width=0.48\textwidth]{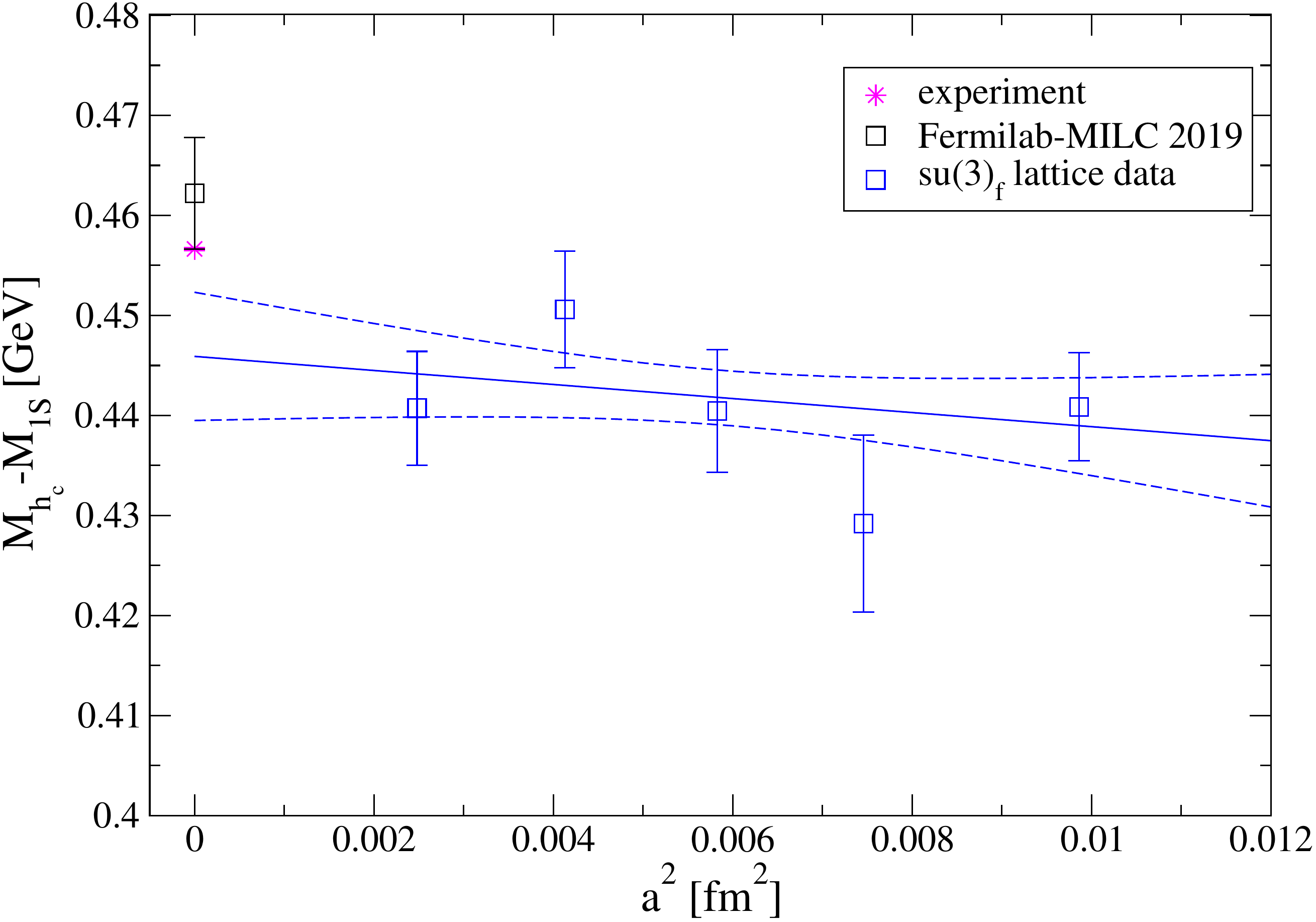}
\caption{Left panel: 1S hyperfine splitting on the trajectory with
  $m_{u/d}=m_s$ corresponding to the heaviest pion masses (choosing the
  largest volume) in Tab.~\ref{configtable}. The splitting is clearly tending
  towards the physical splitting for small lattice spacing. The blue lines and
  blue dashed error band show the results of a fit linear in the lattice
  spacing squared, which is compatible with the presence of such sub-leading
  discretisation effects in our action. Right panel: A proxy for
  the S-wave -- P-wave splitting along the same line. Note that both results are
  not determinations and instead illustrate the quality and suitability of our
tuning.}\label{swavepwave}
\end{figure}

To investigate residual discretisation effects within our approach, we take a
look at the mass-splittings among our tuning states. In the left-hand-side panel of Fig.~\ref{swavepwave} we show the 1S hyperfine splitting between the $J/\psi$ and $\eta_c$
mesons resulting from the neural net training with all ensembles at a given
value of $\beta$. While the run parameters suggested by the neural net based
on the training data do not lead to a particularly accurate 1S hyperfine
splitting at coarse lattice spacing, this discrepancy does get smaller towards
the continuum limit. This suggests that our strategy of using the physical
inputs for the tuning of the action parameters at each lattice spacing is
working. In addition to the data from our final runs we also show the PDG
value \cite{Zyla:2020zbs} as a magenta star, and an independent  lattice
determination that provides this splitting as a prediction \cite{DeTar:2018uko}. For a plot
of further results for this quantity see Fig.~8 of the same reference.

The right hand side of Fig.~\ref{swavepwave} shows a mass-splitting which is
equal to the S-wave -- P-wave splitting in the absence of a P-wave hyperfine
splitting. In this quantity we expect a very mild heavy-quark dependence, as
the physical S-wave -- P-wave splittings in charmonium and bottomonium are
almost the same. Just like for the hyperfine splitting, there is no obvious sign of issues with our tuning approach. Overall we tend to get somewhat smaller values for this splitting, although with large errors coming from our determination of $h_c$. As our
application for this action will mostly be qualitative studies of
hadron-hadron scattering, we currently see no need to attempt to further
improve the observed behaviour, which would require an improvement in statistical precision on our side combined with an improvement in the determination of the lattice spacing.

\subsection{A tuning comparison on the same ensembles}\label{Sec:stupid_tune}

In \cite{Gerardin:2019rua} a charm-quark tuning was performed that uses $c_E=c_B=c_\text{SW}$ and $r_s=\nu=1$, varying $\kappa_c$ to achieve the physical continuum value of the $D_s$-meson mass on the same ensembles as used in this work. This tuning treats the charm-quark on the same footing as the light and strange quarks by using the standard Wilson-clover action, one may well expect this to have significant discretisation effects as $aM_{\eta_c}$ becomes of order 1 for the coarse ensembles.

\begin{figure}[h!]
\includegraphics[width=0.48\textwidth]{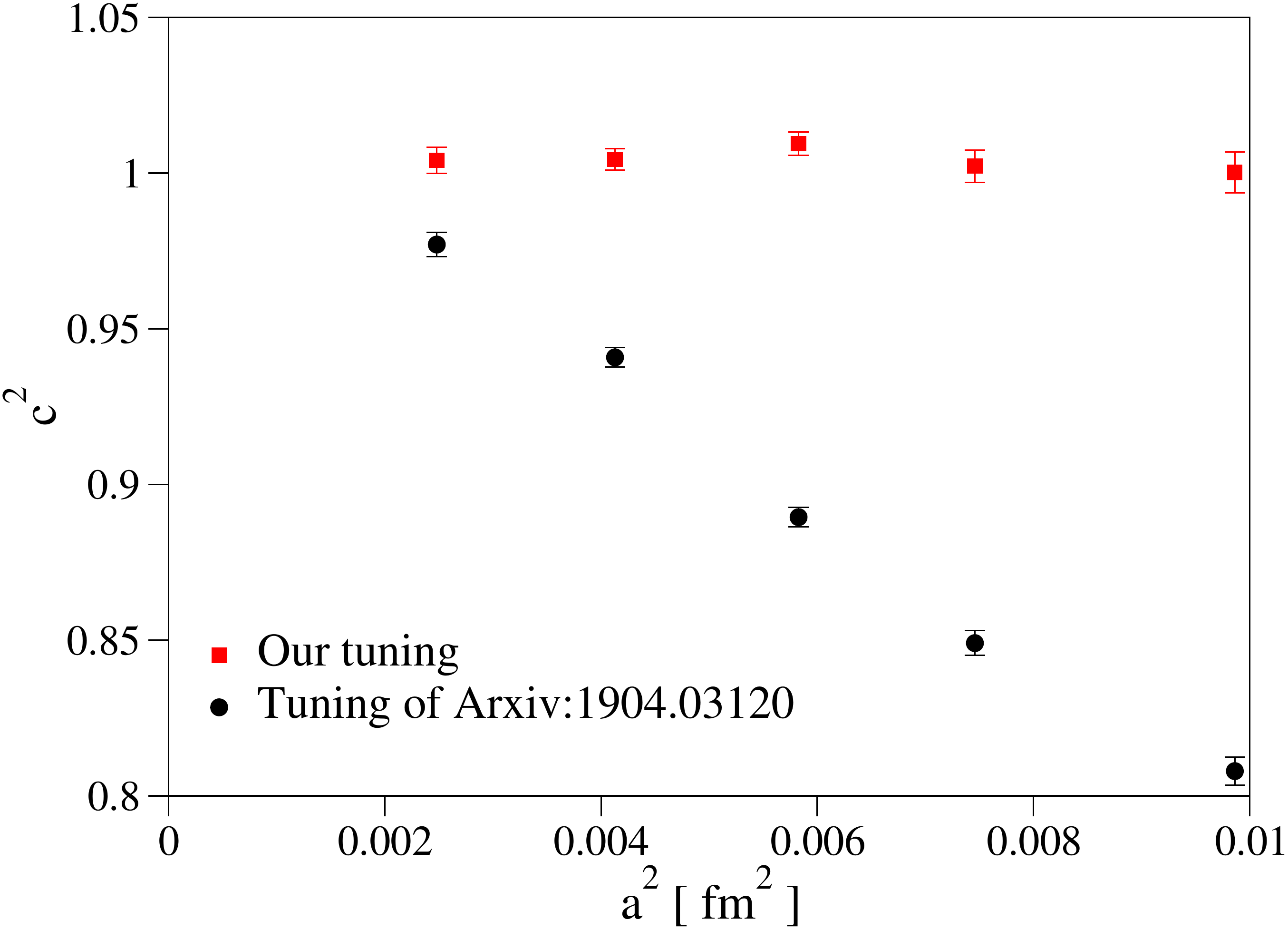}
\includegraphics[width=0.48\textwidth]{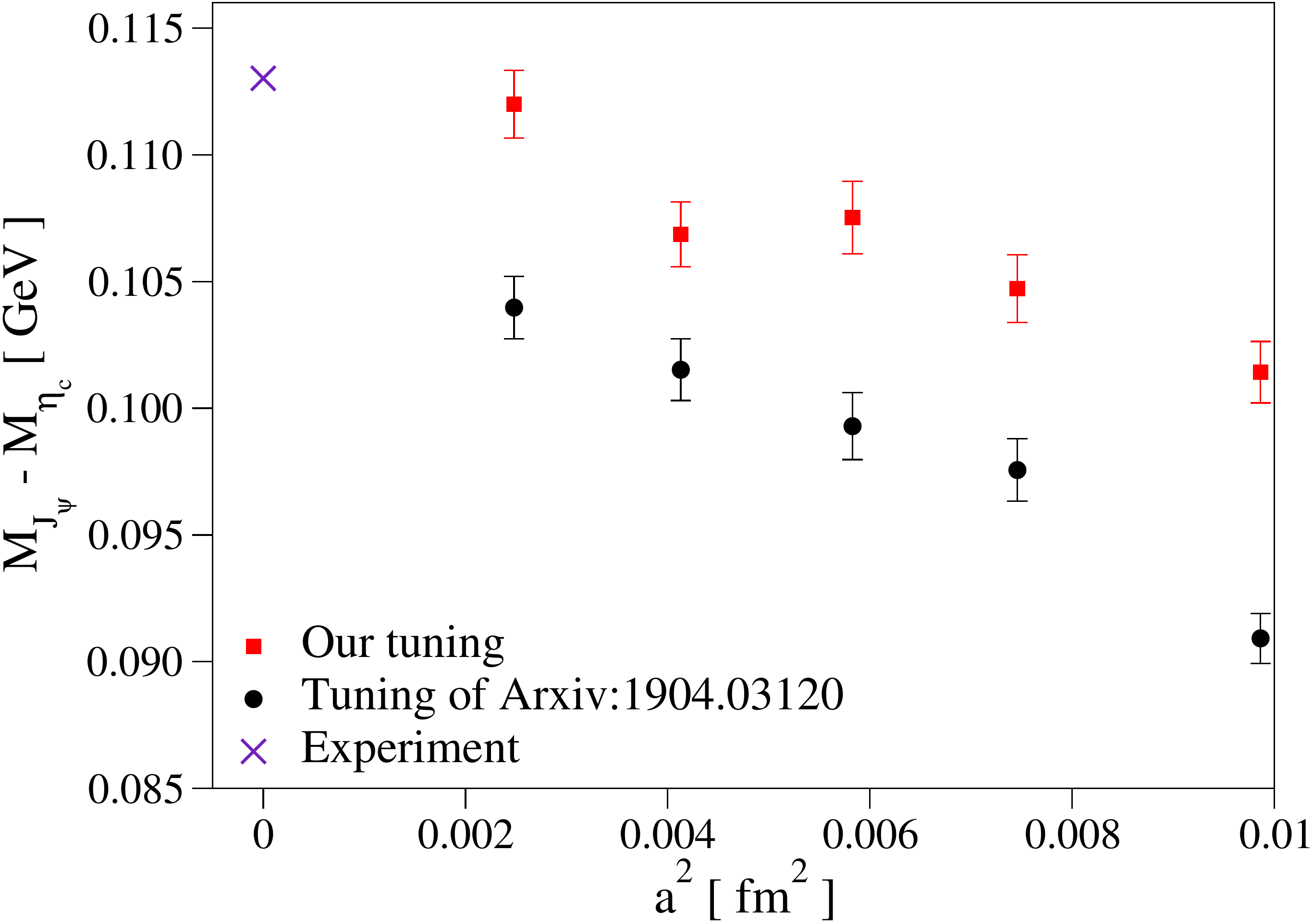}
\caption{(Left) the speed of light squared from the spin-averaged dispersion relation using either the charm-quark tuning of \cite{Gerardin:2019rua} or our tuning on ensembles at the $\text{SU}(3)_f$-symmetric point. (Right) the hyperfine splitting for the two tunings.}\label{fig:comp_ds}
\end{figure}

Fig.~\ref{fig:comp_ds} illustrates the deviation of the spin-averaged
dispersion relation $c^2$, which can be used as a proxy for the magnitude of
discretisation effects of both tunings. In the dispersion relation for the
$D_s$-tuning the deviation from 1 is at most $20\%$ reducing to about $4\%$ on
the finest (N300) lattice. A naive, straight-line fit through all of the
$D_s$-tuned results describes the data poorly and over-predicts the value of
$c^2$ in the continuum as there is still some visible curvature at small
$a^2$. A quadratic fit in $a^2$ to the three finest ensembles describes the
data well and gives $c^2=1$ in continuum.

The hyperfine splitting in Fig.~\ref{fig:comp_ds} shows a similar slope in
$a^2$ between the two approaches although the $D_s$-tuning lies systematically
below that of our tuning and the physical hyperfine splitting. This is due to
the ensembles used in this comparison lying at the $\text{SU}(3)_f$-symmetric
point and the $D_s$-tuning compensating for the valence strange being lighter
than its physical value. This has the side-effect that a precise charm-tuning via the $D_s$ is only valid at the physical point, or through extrapolations toward it when the strange-quark mass is unphysical.

\section{Conclusions and outlook\label{outlook}}

We have shown that it is possible to non-perturbatively determine charm-quark
action parameters reasonably precisely using machine learning techniques. This
method is appealing as it doesn't rely on lattice perturbation theory or an
underlying model to tune parameters. Its main drawback is that a
sufficiently-large number of measurement runs need to be made for training
purposes, which could make it costly for large ensembles. Although, at the
precision we are able to achieve, we see little pion-mass or finite-volume
dependence of our tuning parameters. We also note that it appears impossible
to exactly reproduce the chosen continuum states of charmonium at finite lattice spacing with this 5-parameter action, even though a good approximation can be achieved.

We see no reason why this approach couldn't be extended to, for example,
b-quark physics using the same relativistic heavy-quark action. While residual
discretisation effects are expected to be sizable for bottomonium \cite{Oktay:2008ex},
this approach should be very useful for refining some older predictions for
exotic hadrons in the $B$-meson spectrum \cite{Lang:2015hza}. The same approach could equally-well be used
to tune an effective action, such as NRQCD, provided one has enough continuum states to
compare against that allow for a handle on the underlying
simulation parameters. The method presented here, quite naturally, extends to even higher-order heavy-quark actions due to the inherent flexibility of neural-network fits.

In future works we will use this action for scattering studies using
L\"uscher's finite-volume method for heavy-light meson spectroscopy and
charmonium spectroscopy, with the stochastic distillation technique
\cite{HadronSpectrum:2009krc,Morningstar:2011ka}. In this regard a first step consists of testing the dispersion
relations of heavy-light $D$ and $D_s$ mesons in our approach.

\acknowledgments{The authors would like to acknowledge useful discussions
  with Christopher Johnson and Matthias Lutz and are grateful to Sasa
  Prelovsek for her useful comments on a first draft. D.M. acknowledges funding by the
  Heisenberg Programme of the Deutsche Forschungsgemeinschaft (DFG, German
  Research Foundation) – project number 454605793. R.J.H. by the European Research Council (ERC) under the European
Unions Horizon 2020 research and innovation programme through grant agreement 771971-
SIMDAMA. Calculations for this
  project were partly performed on the HPC cluster “Mogon II” at JGU
  Mainz. This research was supported in part by the cluster computing resource
  provided by the IT Division at the GSI Helmholtzzentrum für
  Schwerionenforschung, Darmstadt, Germany (HPC cluster Virgo). The
  simulations were partly performed on the national supercomputer HPE Apollo Hawk at the High Performance Computing Center Stuttgart (HLRS) under acronym RinQCD4PANDA. For the neural network training and predictions we made use of the Keras API.}

\appendix
\section{On our charm-action implementation}\label{sec:charm_impl}

We use a modified version of the library \verb|OpenQCD-1.6| \cite{Luscher:2012av} for our computation of propagators for the charm-quark Dirace operator of Eq.~\ref{eq:tsuk_action}. This required the rewriting of the Dirac operator in single and double precision in AVX/FMA vector intrinsics as some of the optimisations (and inline assembly) used in the implementation of the \textit{vanilla} Wilson action could not be utilised. On paper the arithmetic intensity of this charm action is about a factor of 2 more expensive than the usual Wilson action. Although an optimisation was used here to turn the underlying spinors into color-major, spin-minor order (\verb|OpenQCD-1.6| is spin-major, color-minor) to improve cache-coherence and perform fewer shuffling and register loading operations. It is likely that this charm-quark action is worse-conditioned than the usual Wilson action as it is found to have a larger clover term than $c_{\text{SW}}$. Nevertheless, our implementation is roughly comparable to a valence strange quark inversion at physical $m_s$ \cite{Hudspith:2020tdf} in \verb|OpenQCD| using the \verb|DFL_SAP_GCR| algorithm \cite{Luscher:2007se}.

In initial investigations we found some deviations between charmonium correlators at large times due to the stopping condition of the $L_2$-norm of the residual depending on the precision we used. To help resolve this we changed the stopping criteria to be the $L_\infty$-norm instead, which empirically behaved much better.

\section{Coulomb gauge-fixed box-sinks as a sink smearing}\label{app:bsink}

If we consider the box-sink propagator introduced in \cite{Hudspith:2020tdf},
\begin{equation}
\tilde{S}(x,t) = \sum_{r=0}^{r^2<R^2} S(x+r,t),
\end{equation}
when one goes to compute the contracted meson there are clearly double-sums and in the limit of $R^2\rightarrow 3\frac{L}{2}^2$ we are performing two sums over the spatial volume, which becomes very expensive. Empirically, we find the cost in contractions scales quadratically with $R^2$, and this sum quickly becomes the dominant cost of contractions as $R^2$ increases.

A more-efficient implementation would be to notice that the box-sink approach could be considered as a convolution with a particular step-function, $f(x) = \theta(r<R^2)$, and herein lies the connection to sink-smearing and a clue to efficient calculation of these propagators. An approximation to this step-function could be a Gaussian with some width $\sigma \propto R^2$,  $f(x) = e^{-\frac{x^2}{\sigma}}$, (although in principle any arbitrary function could be used, for S-wave states something with rotational symmetry makes sense) and we apply the convolution using the usual (fast) Fourier transform prescription,
\begin{equation}
S(p,t) = \mathcal{F}(S(x,t)) = \sum_x e^{ip\cdot x} S(x,t),
\end{equation}
where $\mathcal{F}$ is used to denote the Fourier transform, and likewise $\mathcal{F}^{-1}$ its inverse.
\begin{equation}
\overline{S}(x) = \frac{1}{V}\mathcal{F}^{-1}(S(p,t)f^*(p)).
\end{equation}
We have turned a somewhat expensive approximate convolution into a full convolution with an arbitrary function $f(x)$, the cost of which are $12\times 12$ complex spatial-volume FFTs per time-slice.

It is important to note that this approach only holds because of the Coulomb gauge-fixed wall sources we use, as only in this case can one treat the links as identity matrices and hence completely ignore them and apply a continuous function. Such approaches have been known in the literature in the case of smearing either the source and/or the sink e.g. applying arbitrary potentials in \cite{Davies:1994mp}.

\bibliography{NN}

\end{document}